\documentclass{article}
\usepackage{spconf,amsmath,graphicx}
\usepackage{amsthm,amssymb}

\usepackage{enumitem}
\setlist{nosep, leftmargin=15pt}
\usepackage{mwe} 

\usepackage{hyperref}
\usepackage{listings}
\usepackage{url}
\usepackage{soul}
\usepackage{algorithm}
\usepackage{algorithmic}
\usepackage{caption}
\usepackage{booktabs}
\usepackage{color}
\usepackage{relsize}

\usepackage{multirow}
\usepackage{rotating}

\makeatletter
\def\Hline{%
\noalign{\ifnum0=`}\fi\hrule \@height 1pt \futurelet
\reserved@a\@xhline}
\makeatother

\usepackage{pifont} 

\definecolor{mygreen}{rgb}{0.032, 0.6392, 0.2039}
\newcommand{\cmark}{\textcolor{mygreen}{\ding{51}}}%
\newcommand{\xmark}{\textcolor{red}{\ding{55}}}%
\newcommand{\mymark}{\textcolor{black}}

\title{BAPGAN: GAN-based Bone Age Progression\\ of Femur and Phalange X-ray Images}
\name{Shinji Nakazawa$^1$, Changhee Han$^2$, Joe Hasei$^3$, Ryuichi Nakahara$^4$, Toshifumi Ozaki$^4$ }
\address{$^1$ LPIXEL Inc., Tokyo, Japan\\$^2$ Saitama Prefectural University, Saitama, Japan\\$^3$ Okayama City General Medical Center, Okayama City Hospital, Okayama, Japan \\$^4$ Dept. of Orthopaedic Surgery, Grad. School of Medicine, Dentistry and Pharmaceutical Sciences,\\Okayama University, Okayama, Japan\vspace{-5mm}}

\begin{document}

\maketitle

%

\begin{abstract}

Convolutional Neural Networks play a key role in bone age assessment for investigating endocrinology, genetic, and growth disorders under various modalities and body regions. However, no researcher has tackled bone age progression/regression despite its valuable potential applications: bone-related disease diagnosis, clinical knowledge acquisition, and museum education. Therefore, we propose Bone Age Progression Generative Adversarial Network (BAPGAN) to progress/regress both femur/phalange X-ray images while preserving identity and realism. We exhaustively confirm the BAPGAN's clinical potential $via$ Fr\'{e}chet Inception Distance, Visual Turing Test by two expert \mymark{
orthopedists}, and t-Distributed Stochastic Neighbor Embedding.

\end{abstract}
\begin{keywords}
G\hspace{-0.8mm}enerative Adversarial Networks, Medical\\ Image Synthesis, Bone X-ray, Age Progression, Visual Turing Test
\end{keywords}

\vspace{-7mm}

\section{Introduction}
\vspace{-2mm}
\label{sec:intro}
Because skeletal maturity progresses through discrete stages, pediatric medicine has correlated children's chronological age with bone age to investigate endocrinology, genetic, and growth disorders; but, time-consuming manual bone age assessment methods~\cite{greulich1959radiographic, tanner2001assessment} suffer from intra- and inter-observer variability.  In this context, Convolutional Neural Networks have shown great promise in age assessment on various modalities and body regions, including hand/pelvic X-ray~\cite{spampinato2017deep, fan2020evaluation}, clavicula Computed Tomography~\cite{gassenmaier2020forensic}, and hand Magnetic Resonance \mymark{Imaging}~\cite{vstern2016automated}.


Along with the assessment, bone age progression/regression \mymark{(i.e., predicting a given bone image's future/past appearance)} also matters because obtaining desired bone age images could lead to valuable applications: diagnosis (e.g., locating tumors around growth plates, such as osteosarcoma and aneurysmal bone cysts); clinical knowledge acquisition (e.g., how progression in epiphyseal width relates to epiphyseal fusion); museum education \mymark{for children} (e.g., height prediction). In Computer Vision, such face age progression using Generative Adversarial Networks (GANs) plays a key role in age-invariant verification/entertainment; Conditional Adversarial Auto\mymark{E}ncoder (CAAE) is known for its age progression without paired samples~\cite{CAAE}. However, no medical imaging researcher has tackled this despite the GANs' growing clinical attention~\cite{yu2019ea, han2019combining}. Moreover, no bone age assessment method has analyzed multiple datasets~\cite{10.1371/journal.pone.0220242}.


\begin{figure}[t!]
  \centering
  \centerline{\includegraphics[width=1\columnwidth]{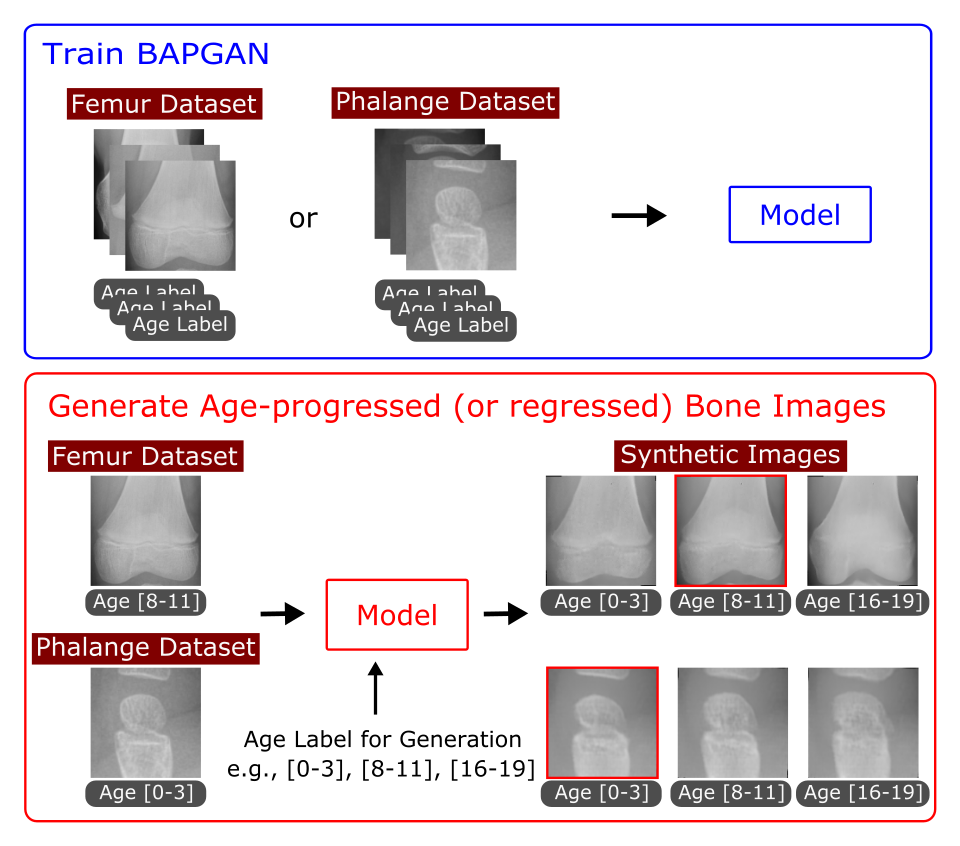}}
\caption{BAPGAN-based identity-preserved and realistic \mymark{$128 \times 128$} X-ray image generation at desired bone age.}

\label{fig:bone_age_progression}
\vspace{-5mm}
\end{figure}

We propose Bone Age Progression GAN (BAPGAN) to progress/regress both femur/phalange X-ray images while preserving identity and realism (Fig.~\ref{fig:bone_age_progression}). \mymark{Since our task targeting ages $0$--$19$ requires capturing short-term local subtle changes (especially around joints/growth plates) unlike the face age progression targeting all ages, we make these modifications to the CAAE: (\textit{i}) an age discriminator; (\textit{ii}) age label smoothing; (\textit{iii}) Self-Attention (SA) modules.} We exhaustively confirm its clinical potential $via$ Fr\'{e}chet Inception Distance (FID)~\cite{heusel2017gans}, Visual Turing Test~\cite{Salimans} by two expert \mymark{orthopedists}, and t-Distributed Stochastic Neighbor Embedding (t-SNE)~\cite{Maaten}.

\begin{figure*}[t!]
    \centering
    \includegraphics[width=2\columnwidth]{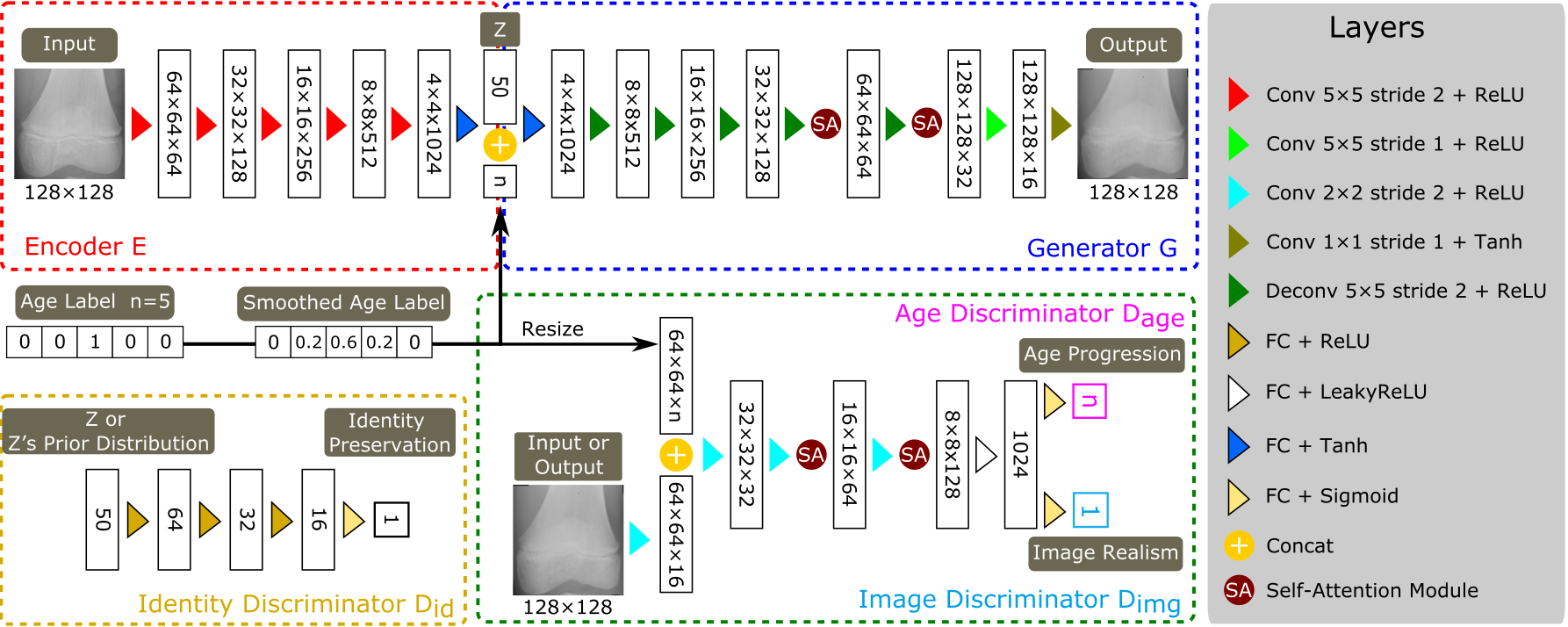}
    \caption{BAPGAN architecture for identity-preserved, realistic, and age-progressed $128 \times 128$ bone \mymark{image} generation: (\textit{a}) the identity discriminator $D_{id}$ aims to classify a latent variable $z$ \textit{vs} its prior distribution; (\textit{b}) the image discriminator $D_{img}$ learns to classify real \textit{vs} synthetic bones; (\textit{c}) the age discriminator $D_{age}$ learns to categorize bone age.}
    \label{fig:structure}
    \vspace{-3mm}
\end{figure*}

\noindent \textbf{Contributions.} Our main contributions are as follows:
\begin{itemize}
\item \textbf{Multi-Dataset Bone Age Analysis:} This computational study firstly investigates how joint/growth plate appearance\mymark{s are} associated with bone age on multiple datasets.

\item \textbf{Bone Age Progression:} This first bone age progression approach shows that BAPGAN can generate identity-preserved/realistic X-ray images at desired bone age.

\item \textbf{Clinical Potential:} This preliminary research qualitatively and quantitatively assesses BAPGAN-generated images for potential valuable applications: bone-related disease diagnosis, clinical knowledge acquisition, and museum education.
\end{itemize}


\vspace{-3mm}

\section{Materials and Methods}
\label{sec:Materials and Methods}

\subsection{Femur X-ray Dataset}
We use a dataset of the  lower extremity of femur X-ray images, cropped from healthy left leg X-ray images (image resolution: $1,793\pm{975} \times 1,718\pm{428}$). The dataset was collected by the authors at Suiwakai Mizushima Central Hospital, Okayama, Japan, which is currently not publicly available due to ethical restrictions. The images are resized to $128 \times 128$ pixels after rotation correction. Subjects' age intervals are defined as: age [$0$-$3$]/[$4$-$7$]/[$8$-$11$]/[$12$-$15$]/[$16$-$19$] ($42$
/$47$/$56$/$130$/$42$ images, respectively).

Similar to height increase, growth plates in children's femur stop growing and close near the end of puberty through discrete stages; similarly, surrounding bones gradually stop connecting to their joints. Thus, we use a 4-year equal interval classification scheme for bone age progression/regression. The dataset \mymark{with $317$ images} is randomly split into: training/validation/test sets ($223$/$32$/$62$ images, respectively).








\subsection{Phalange X-ray Dataset}
\vspace{-1.5mm}
This paper also uses a dataset of the middle phalange of index finger X-ray images, cropped from healthy left hand X-ray images (image resolution: $1,337\pm{238} \times 1,680\pm{300}$). The dataset was extracted from RSNA Pediatric Bone Age Challenge~\cite{RSNA}. The images are resized to $128 \times 128$ pixels after rotation correction. Subjects' age intervals are defined as: age [$0$-$3$]/[$4$-$7$]/[$8$-$11$]/[$12$-$15$]/[$16$-$19$] ($137$/$1,309$
/$2,514$/$2,912$/$293$ images, respectively).

The phalange dataset is imbalanced towards age [$4$-$15$] as the challenge \mymark{mainly predicts} the age of middle childhood/early adolescence. The growth of children's phalange is smaller than that of children's femur on the images. The dataset \mymark{with $7,165$ images} is randomly split into: training/validation/test sets ($5,015$/$717$/$1,433$ images, respectively).

\vspace{-3mm}





\vspace{-2mm}
\subsection{Proposed Bone Age Progression Approach}
\vspace{-2mm}
\subsubsection{BAPGAN-based Bone Age Progression}
\vspace{-2.2mm}

\noindent \textbf{BAPGAN} is a novel image-to-image GAN for bone age progression/regression (Fig. \ref{fig:structure}). \mymark{It extends CAAE~\cite{CAAE}, which was originally conceived for face age progression/regression. The BAPGAN aims to capture short-term local subtle bone appearance changes between ages $0$-$19$.}

\mymark{The CAAE, which is a} GAN and AAE-combined end-to-end architecture, consists of four networks: an encoder \(E\), a generator \(G\), an identity discriminator \(D_{id}\), and an image discriminator \(D_{img}\). \(E\) encodes an input image \(x\) to a latent variable \(z\) to preserve the subject's identity. \(G\) generates a synthetic image \(x'\) given \(z\) and the subject's \mymark{one-hot vector} age label \(l\). \mymark{\(D_{id}\) aims to classify $z$ \textit{vs} its prior uniform distribution \(p(z)\) for identity-preserved image generation while \(D_{img}\) learns to classify \(x\) \textit{vs} \(x'\) for realistic/plausible image generation.} To sum up, \mymark{the} CAAE \mymark{(a version without a total variation loss)} adopts the following loss:
\vspace{-3mm}

\footnotesize
\begin{equation}
\begin{aligned}
\mathcal{L}_{CAAE}\ (x,l) &= \min_{E,G}\max_{D_{id}, D_{img}} \lambda_1\mathcal{L} (x, G(E(x),l))\\
&+\mymark{\mathbb{E}}_{z^\star\sim p(z)} \bigl[\log D_{id}(z^\star)\bigr] \\
&+\mymark{\mathbb{E}}_{x\sim p_{d}(x)}\bigl[\log (1-D_{id}(E(x)))\bigr] \\
&+\mymark{\mathbb{E}}_{x,l\sim p_{d}(x,l)}\bigl[\log D_{img}(x,l)\bigr] \\
&+\mymark{\mathbb{E}}_{x,l\sim p_{d}(x,l)}\bigl[\log(1-D_{img}(G(E(x),l)))\bigr],
\end{aligned}
\end{equation}
\normalsize

 where \(p_{d}(x,l)\) denotes a training data distribution, \(z^{\star}\) \mymark{indicates} \(z\) sampled from the prior, and \(\mathcal{L} \) \mymark{is} $\ell _2$ norm. \(\lambda_1\) \mymark{balances smoothness and high resolution \cite{CAAE}}.

As shown in Fig.~\ref{fig:structure}, \mymark{our} BAPGAN adopts the following modifications to improve realism and age progression:

\vspace{1mm}
\noindent \textbf{Age Discriminator} \mymark{To explicitly represent aging effects}, we introduce an age discriminator $D_{age}$ and employ multi-task learning with \(D_{img}\). \mymark{The loss from classifying \(x\) into $6$ age classes helps optimize \(D_{img}\) while the loss from classifying \(x'\) into $6$ age classes helps optimize \(G\) as follows:}
\footnotesize
\begin{equation}
\begin{aligned}
\mathcal{L}_{D_{age}}\ (x,l) &= \min_{E,G}\max_{D_{age}} \mymark{\mathbb{E}}_{x,l\sim p_{d}(x,l)}\bigl[-\log D_{age}(l|x)\bigr] \\
 &+  \mymark{\mathbb{E}}_{x,l\sim p_{d}(x,l)}\bigl[-\log D_{age}(l|G(E(x),l))\bigr]. \end{aligned}
\end{equation}
\normalsize

\vspace{1mm}
\noindent \textbf{Age Label Smoothing} To regularize ambiguous adjacent bone age representation, we smooth \(l\) with smoothing $0.2$ (e.g., [$0$, $0$, $1$, $0$, $0$] to [$0$, $0.2$, $0.6$, $0.2$, $0$], [$1$, $0$, $0$, $0$, $0$] to [$0.8$, $0.2$, $0$, $0$, $0$]). \(l^{LS}\) denotes such a smoothed age label.


\vspace{1mm}
\noindent \textbf{SA Modules} 
To ignore noise and concentrate on bone age-relevant body parts by learning global and long-range dependencies, we apply SA modules after the designated convolutional/deconvolutional layers in $G$/$D_{img}$/$D_{age}$. Following the original paper~\cite{SAGAN}, we also adopt spectral normalization~\cite{Spectral} after each convolutional layer in $D_{id}$/$D_{img}$/$D_{age}$ and imbalanced learning rate for stable GAN training.


\vspace{1mm}

Finally, our BAPGAN loss becomes:
\footnotesize
\begin{eqnarray}
\mathcal{L}_{BAPGAN}\ (x,l^{LS}) = \mathcal{L}_{CAAE}\ (x,l^{LS}) + \mymark{\lambda_2}\mathcal{L}_{D_{age}}\ (x,l^{LS}),
 \end{eqnarray}
\normalsize
where \mymark{\(\lambda_2\)} is a weight coefficient of $\mathcal{L}_{D_{age}}$.

\vspace{2mm}
\noindent \textbf{BAPGAN Implementation Details}
We compare BAPGAN against CAAE, CAAE with only an age discriminator/age label smoothing, and CAAE with only SA modules (and stabilization techniques). Each GAN training lasts for $5.0 \times 10^{4}$ steps with a batch size of $32$. With Adam optimizer (\(\beta_1=0.5, \beta_2=0.999\)), we use $1.0 \times 10^{-4}$ learning rate for \(E\)/\(G\)/\(D_{id}\) and $4.0 \times 10^{-4}$ for \(D_{img} / D_{age}\). The prior is a uniform distribution with $50$ dimension. \mymark{$\lambda_1$ and $\lambda_2$ are set to be $10000$ and $100$.} As data augmentation, after random 5-degree rotation, we apply randomly-selected two color enhancing operations (i.e., \mymark{a} limited version of RandAugment~\cite{RandAug}) from [\texttt{auto-contrast}, \texttt{contrast},\\ \texttt{brightness}, \texttt{sharpness}, \texttt{posterize}].



\begin{figure}[!t]
\vspace{-2mm}
  \centering
  \centerline{\includegraphics[width=0.94\linewidth]{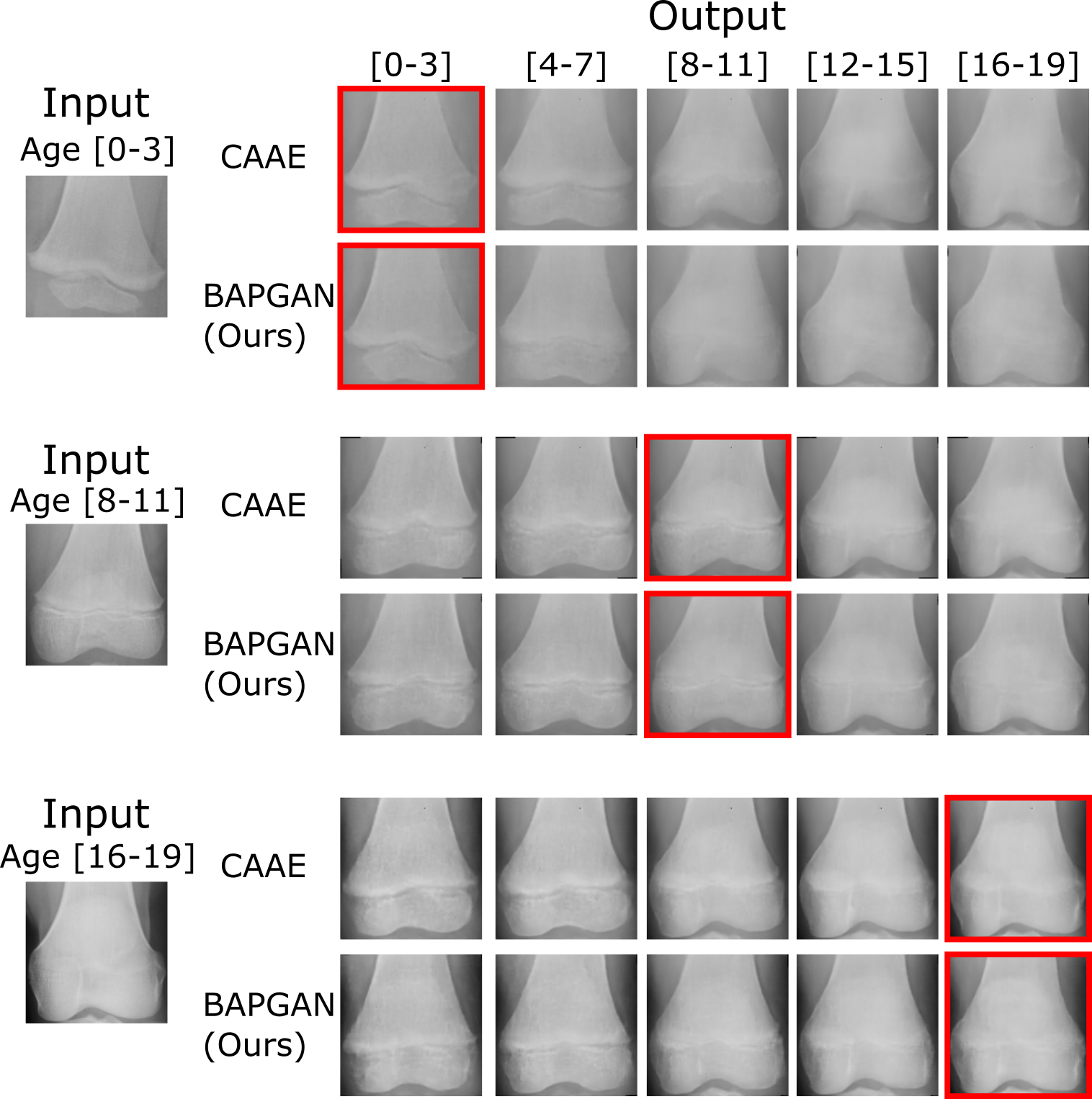}}
 \vspace{-2mm}
\caption{Example synthetic $128 \times 128$ Femur images. Red boxes denote age-invariant reconstruction results.}
\label{fig:generated-femur}
\vspace{-6mm}
\end{figure}

\vspace{-4mm}
\subsubsection{Clinical Validation Using Visual Turing Test}
\vspace{-1.5mm}
On both femur/phalange datasets, we evaluate the (\textit{i}) realism of $128 \times 128$ synthetic images by CAAE and BAPGAN respectively against real images and its (\textit{ii}) age progression/regression performance. First, we supply, in random order, to two \mymark{orthopedists} a random selection of $50$ real and $50$ synthetic images. Then, the \mymark{orthopedists} are asked to classify them as real/synthetic. Similarly, the \mymark{orthopedists} also have to select age-progressed images between $25$ real vs $25$ their 8-year CAAE/BAPGAN-progressed images. We perform the same for $8$-year regressed ones. \mymark{It should be noted that we use a relatively-long age interval (i.e., 8-year) because of the limited clinical data against the subtle appearance changes.} The Visual Turing Test~\cite{Salimans} can probe physicians' ability to identify clinical attributes in GAN-generated images~\cite{han2018gan}.


%

%

\vspace{-4mm}
\subsubsection{Visualization Using t-SNE}
\vspace{-1.5mm}
T-SNE~\cite{Maaten} visualizes inter-age distributions of real/synthetic $128 \times 128$ femur/phalange images, respectively, on a random selection of $210$ real, $210$ CAAE-generated, and $210$ BAPGAN-generated images. We select $42$ images per each age label ($210$ in total); $42$ is the maximum available number of the femur real images for age [0-3].

\vspace{1mm}
\noindent \textbf{T-SNE Implementation Details}
\mymark{T-SNE training lasts for $500$ steps with $50$ perplexity. We normalize pixels to $[0, 1]$.}



\setcounter{table}{1}

\begin{table*}[!t]
 \caption{Visual Turing Test results by two \mymark{orthopedists} for classifying $50$ real (\textbf{R}) vs $50$ synthetic (\textbf{S}) images by CAAE and BAPGAN respectively. Proximity to $50\%$ of accuracy is superior (chance = $50\%$).}
 \label{table:realism}
 \centering
  \begin{tabular}{p{0.3em}ccccccc}
    \Hline\noalign{\smallskip}
&     & \bfseries Orthopedist & \bfseries Accuracy & \bfseries R as R & \bfseries R as S & \bfseries S as R & \bfseries S as S \vspace{0.7mm} \\
\hline\noalign{\smallskip}
\parbox[t]{0mm}{\multirow{4}{*}{\rotatebox[origin=c]{270}{\textbf{\shortstack{CAAE}}}}} &
   \parbox[t]{24mm}{\multirow{2}{*}{\bfseries Femur Dataset}} & A &\textbf{67}\% & 74\% & 26\% & 43\% & 57\% \\
      &  & B &\textbf{74}\% & 95\% & 5\% & 55\% & 45\% \\
&  \parbox[t]{24mm}{\multirow{2}{*}{\bfseries Phalange Dataset}} & A &73\% & 83\% & 17\% & 41\% & 59\% \\
&  & B &85\% & 83\% & 17\% & 12\% & 88\% \\
    \hline\noalign{\smallskip}
\parbox[t]{0mm}{\multirow{4}{*}{\rotatebox[origin=c]{270}{\textbf{\shortstack{BAPGAN}}}}} &
   \parbox[t]{24mm}{\multirow{2}{*}{\bfseries Femur Dataset}} & A &76\% & 90\% & 10\% & 43\% & 57\% \\
      &  & B &86\% & 97\% & 3\% & 29\% & 71\% \\
&  \parbox[t]{24mm}{\multirow{2}{*}{\bfseries Phalange Dataset}} & A &\textbf{66}\% & 69\% & 31\% & 39\% & 61\% \\
&  & B &\textbf{66}\% & 90\% & 10\% & 68\% & 32\% \\
\Hline\noalign{\smallskip}
  \end{tabular}
\vspace{-2mm}
\end{table*}

\begin{table*}[!t]
 \caption{Visual Turing Test results by \mymark{orthopedists} for selecting age-progressed images between $25$ real vs $25$ their $8$-year CAAE/BAPGAN-progressed images. We perform the same for $8$-year age-regressed ones. Proximity to $100\%$ of accuracy is superior (chance = $50\%$).}
 \label{table:age_progression}
 \centering
  \begin{tabular}{p{0.3em}ccccc}
    \Hline\noalign{\smallskip}
&     & \bfseries Orthopedist & \bfseries Accuracy & \bfseries Progression & \bfseries Regression \vspace{0.7mm} \\
\hline\noalign{\smallskip}
\parbox[t]{0mm}{\multirow{4}{*}{\rotatebox[origin=c]{270}{\textbf{\shortstack{CAAE}}}}} &
   \parbox[t]{24mm}{\multirow{2}{*}{\bfseries Femur Dataset}} & A & 94\% & 100\% & 88\%\\
 & & B & \textbf{96}\% & 100\% & 92\%\\
 & \parbox[t]{24mm}{\multirow{2}{*}{\bfseries Phalange Dataset}} & A & \textbf{48}\% & 88\% & 8\%\\
 & & B & 54\% & 40\% & 68\%\\
\hline\noalign{\smallskip}
\parbox[t]{0mm}{\multirow{4}{*}{\rotatebox[origin=c]{270}{\textbf{\shortstack{BAPGAN}}}}} &
   \parbox[t]{24mm}{\multirow{2}{*}{\bfseries Femur Dataset}} & A & \textbf{100}\% & 100\% & 100\%\\
 & & B & \textbf{96}\% & 100\% & 92\%\\
 & \parbox[t]{24mm}{\multirow{2}{*}{\bfseries Phalange Dataset}} & A & 46\% & 60\% & 32\%\\
 & & B & \textbf{60}\% & 72\% & 48\%\\
\Hline\noalign{\smallskip}
  \end{tabular}
 \vspace{-2mm}
  \end{table*}

\vspace{-4mm}
\section{Results}
\label{sec:results}
\vspace{-2.5mm}

\subsection{BAPGAN-generated \mymark{Femur/Phalange} X-ray Images}
\begin{figure}[!t]
  \centering
  \vspace{-2mm}
  \centerline{\includegraphics[width=0.94\linewidth]{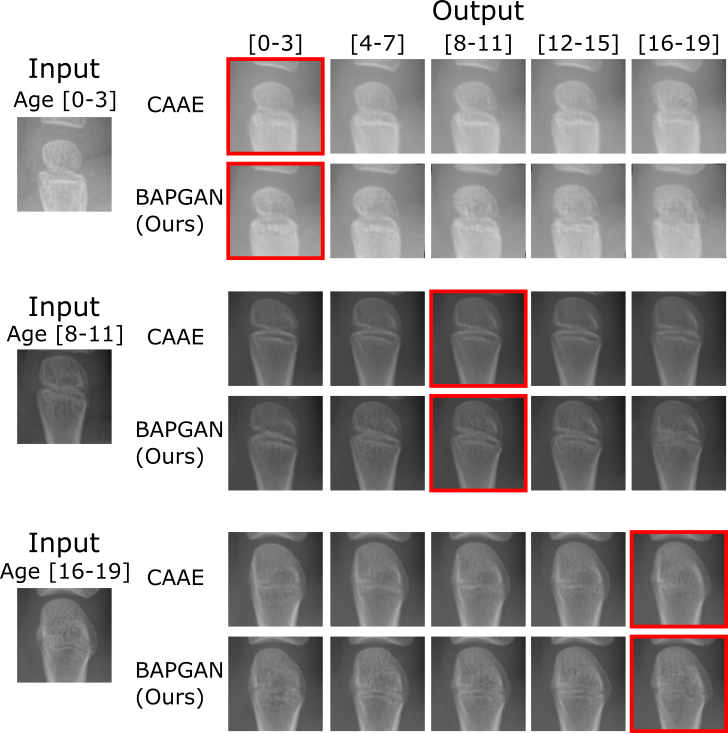}}
 \vspace{-2mm}
\caption{Example synthetic $128 \times 128$ Phalange images. Red boxes denote age-invariant reconstruction results.}
\label{fig:generated-phalange}
\vspace{-2mm}
\end{figure}

Our BAPGAN progresses/regresses bone images more clearly than CAAE, especially around joints and growth plates (Figures~\ref{fig:generated-femur} and~\ref{fig:generated-phalange}). Moreover, as measured by FID \mymark{(calculated between real images/their age-invariant reconstruction over the whole data distribution using Inception-v3~\cite{szegedy2016rethinking})}, the BAPGAN can generate \mymark{remarkably} more realistic images \mymark{(Table~\ref{table:FIDscore})}. A femur dataset shows comparatively high FID, especially with SA modules, due to its limited training data.

\setcounter{table}{0}

\begin{table}[!t]
 \caption{FID for ablations of our proposed modifications to CAAE: $D_{age}$, LS, and SA represent an age discriminator, label smoothing, and SA modules, respectively.}
 \label{table:FIDscore}
 \centering
\begin{tabular}{p{4em}ccccc}
\Hline\noalign{\smallskip}
 & \bfseries Model \vspace{-0.5mm} & \bfseries \hspace{-3mm} D$_{age}$  & \bfseries LS & \bfseries SA  & \bfseries FID \\\noalign{\smallskip}\hline\noalign{\smallskip}
\parbox[t]{4mm}{\multirow{4}{*}{\textbf{\shortstack{\hspace{1mm}Femur\\\hspace{1mm}Dataset}}}} & CAAE~\cite{CAAE} & \hspace{-3mm} \xmark & \xmark & \xmark & 125.82\\ & + D$_{age}$, LS \hspace{1.8mm}\ \ \  & \hspace{-3mm} \cmark & \cmark & \xmark  & 121.12\\
& + SA \hspace{11.693mm} & \hspace{-3mm} \xmark  & \xmark  & \cmark  & 148.13\\
& \hspace{-1mm} BAPGAN & \hspace{-3mm} \cmark  & \cmark  & \cmark  & \textbf{102.53}\\
\noalign{\smallskip}\hline\noalign{\smallskip}
\parbox[t]{4mm}{\multirow{4}{*}{\textbf{\shortstack{Phalange\\Dataset}}}} & CAAE~\cite{CAAE} &  \hspace{-3mm} \xmark & \xmark & \xmark & 94.98\\ & + D$_{age}$, LS \hspace{1.8mm}\ \ \  & \hspace{-3mm} \cmark & \cmark & \xmark  & 76.25\\ & + SA \hspace{11.693mm} & \hspace{-3mm} \xmark  & \xmark  & \cmark  & 79.64\\ & \hspace{-1mm} BAPGAN & \hspace{-3mm} \cmark  & \cmark  & \cmark  & \textbf{43.73}\\
\noalign{\smallskip}\Hline\noalign{\smallskip}
\end{tabular}
\vspace{-2mm}
\end{table}

\subsection{Visual Turing Test Results}
\label{ssec:visual_turing_test}
Two expert \mymark{orthopedists} fail to accurately recognize BAPGAN-generated images thanks to their realism and succeeds to recognize BAPGAN-progressed/regressed femur images (Tables~\ref{table:realism} and~\ref{table:age_progression}). Meanwhile, recognizing phalange progression/regression is challenging due to its smaller growth on the images and training data imbalance.







\begin{figure*}[!t]
\begin{minipage}[b]{1.0\linewidth}
  \centering
  \centerline{\includegraphics[width=\textwidth]{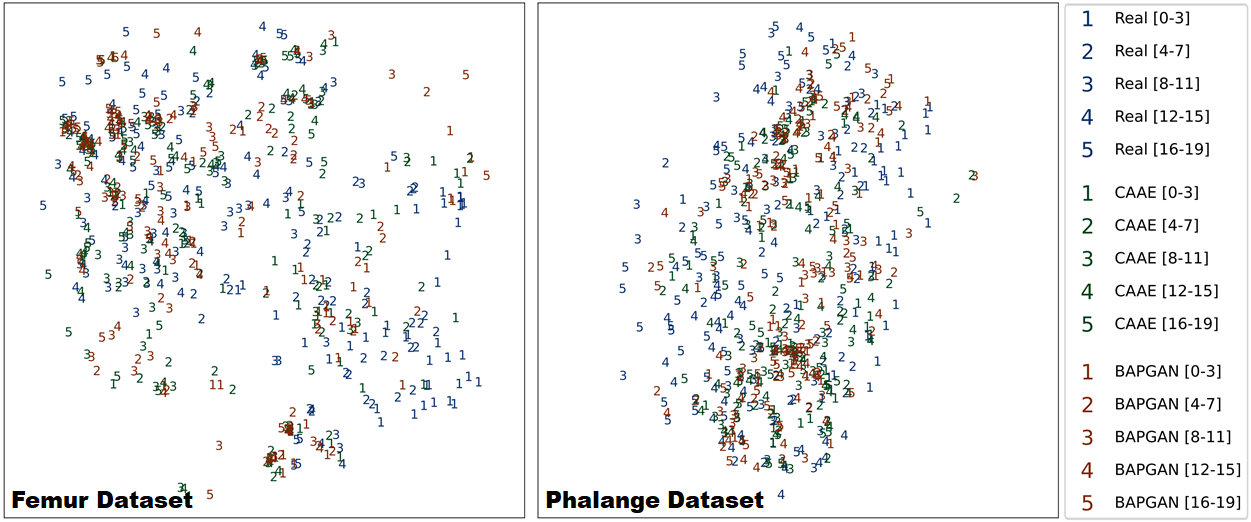}}
\end{minipage}
\caption{T-SNE plots with bone images: (a) real, (b) CAAE-generated, (c) BAPGAN-generated bones.}
\label{fig:tsne}
\vspace{-6mm}
\end{figure*}  
 
\subsection{T-SNE Results}
 \vspace{-1mm}
\label{ssec:t-SNE}
Real femur images for each age category are separately distributed except for age [12-15] and [16-19] since bone growth becomes slower until it ceases altogether (Figure~\ref{fig:tsne}). Meanwhile, distributions of real phalange images for age [4-19] highly overlap due to its smaller growth. BAPGAN-generated images have similar distributions to the real ones.




\vspace{-4mm}

\section{CONCLUSION}
 \vspace{-2mm}
\label{sec:conclusiont}
As confirmed by two expert \mymark{orthopedists}, our BAPGAN can translate both femur/phalange X-ray images into identity-preserved and realistic ones at desired bone age---those generated images could lead to valuable clinical applications (i.e., bone-related disease diagnosis, clinical knowledge acquisition, and museum education). This attributes to the BAPGAN's good realism/age-progression performance capturing subtle \mymark{bone appearance} changes by combining an age discriminator/label smoothing with SA modules. Our BAPGAN points out new directions for research on time-series fine-grained image analysis in Computer Vision, such as remote sensing change prediction.

Our future work include considering sex differences since boys/girls have different timing of puberty (i.e., bone maturation); height growth is over when the boys/girls reach bone age 18/16, respectively~\cite{parthasarathy2019iap}. \mymark{Moreover, to synthesize and evaluate images with exact bone age, we plan to collect more training data and add class weights for handling class imbalance.} Considering all distal/middle/proximal phalanges might be also helpful. Finally, We will investigate more GAN stabilization techniques, such as GAN losses~\cite{mao2017least}, $\ell _1$/$\ell _2$ norm combination, and truncation trick~\cite{brock2018large}.








\bibliographystyle{IEEEbib}
\bibliography{strings,refs}

\begin{thebibliography}{10}

\bibitem{greulich1959radiographic}
W.~W. Greulich and S.~I. Pyle,
\newblock {\em Radiographic atlas of skeletal development of the hand and
  wrist},
\newblock Stanford university press, 1959.

\bibitem{tanner2001assessment}
J.~M. Tanner, R.~H. Whitehouse, N.~Cameron, et~al.,
\newblock {\em Assessment of skeletal maturity and prediction of adult height
  ({TW2} method)},
\newblock Saunders London, 2001.

\bibitem{spampinato2017deep}
C.~Spampinato, S.~Palazzo, D.~Giordano, et~al.,
\newblock ``Deep learning for automated skeletal bone age assessment in {X}-ray
  images,''
\newblock {\em Med. Image Anal.}, vol. 36, pp. 41--51, 2017.

\bibitem{fan2020evaluation}
F.~Fan, X.~Dong, X.~Wu, et~al.,
\newblock ``An evaluation of statistical models for age estimation and the
  assessment of the 18-year threshold using conventional pelvic radiographs,''
\newblock {\em Forensic Sci. Int.}, vol. 314, pp. 110350, 2020.

\bibitem{gassenmaier2020forensic}
S.~Gassenmaier, J.~F. Schaefer, K.~Nikolaou, et~al.,
\newblock ``Forensic age estimation in living adolescents with {CT} imaging of
  the clavicula—impact of low-dose scanning on readers’ confidence,''
\newblock {\em Eur. Radiol.}, pp. 1--8, 2020.

\bibitem{vstern2016automated}
D.~{\v{S}}tern, C.~Payer, V.~Lepetit, and M.~Urschler,
\newblock ``Automated age estimation from hand {MRI} volumes using deep
  learning,''
\newblock in {\em Proc. International Conference on Medical Image Computing and
  Computer-Assisted Intervention (MICCAI)}. Springer, 2016, pp. 194--202.

\bibitem{CAAE}
Z.~Zhang, Y.~Song, and H.~Qi,
\newblock ``Age progression/regression by conditional adversarial
  autoencoder,''
\newblock in {\em Proc. IEEE conference on Computer Vision and Pattern
  Recognition (CVPR)}, 2017, pp. 5810--5818.

\bibitem{yu2019ea}
B.~Yu, L.~Zhou, L.~Wang, et~al.,
\newblock ``{EA-GAN}s: edge-aware generative adversarial networks for
  cross-modality {MR} image synthesis,''
\newblock {\em IEEE Trans. Med. Imaging}, vol. 38, no. 7, pp. 1750--1762, 2019.

\bibitem{han2019combining}
C.~Han, L.~Rundo, R.~Araki, et~al.,
\newblock ``Combining noise-to-image and image-to-image {GAN}s: brain {MR}
  image augmentation for tumor detection,''
\newblock {\em IEEE Access}, vol. 7, pp. 156966--156977, 2019.

\bibitem{10.1371/journal.pone.0220242}
A.~L. Dallora, P.~Anderberg, O.~Kvist, et~al.,
\newblock ``Bone age assessment with various machine learning techniques: a
  systematic literature review and meta-analysis,''
\newblock {\em PloS one}, vol. 14, no. 7, pp. e0220242, 2019.

\bibitem{heusel2017gans}
M.~Heusel, H.~Ramsauer, T.~Unterthiner, et~al.,
\newblock ``{GAN}s trained by a two time-scale update rule converge to a local
  nash equilibrium,''
\newblock in {\em Advances in Neural Information Processing Systems (NIPS)},
  2017, pp. 6626--6637.

\bibitem{Salimans}
T.~Salimans, I.~Goodfellow, W.~Zaremba, et~al.,
\newblock ``Improved techniques for training {GAN}s,''
\newblock in {\em Advances in Neural Information Processing Systems (NIPS)},
  2016, pp. 2234--2242.

\bibitem{Maaten}
L.~van~der Maaten and G.~Hinton,
\newblock ``Visualizing data using {t-SNE},''
\newblock {\em J. Mach. Learn. Res.}, vol. 9, pp. 2579--2605, 2008.

\bibitem{RSNA}
S.~S. Halabi, L.~M. Prevedello, J.~Kalpathy-Cramer, et~al.,
\newblock ``The {RSNA} pediatric bone age machine learning challenge,''
\newblock {\em Radiology}, vol. 290, no. 2, pp. 498--503, 2019.

\bibitem{SAGAN}
H.~Zhang, I.~Goodfellow, D.~Metaxas, and A.~Odena,
\newblock ``Self-attention generative adversarial networks,''
\newblock in {\em Proc. International Conference on Machine Learning (ICML)}.
  PMLR, 2019, pp. 7354--7363.

\bibitem{Spectral}
T.~Miyato, T.~Kataoka, M.~Koyama, and Y.~Yoshida,
\newblock ``Spectral normalization for generative adversarial networks,''
\newblock in {\em Proc. International Conference on Learning Representations
  (ICLR) arXiv:1802.05957}, 2018.

\bibitem{RandAug}
E.~D. Cubuk, B.~Zoph, J.~Shlens, and Q.~V. Le,
\newblock ``Randaugment: practical automated data augmentation with a reduced
  search space,''
\newblock in {\em Proc. IEEE/CVF Conference on Computer Vision and Pattern
  Recognition (CVPR) Workshops}, 2020, pp. 702--703.

\bibitem{han2018gan}
C.~Han, H.~Hayashi, L.~Rundo, et~al.,
\newblock ``{GAN}-based synthetic brain {MR} image generation,''
\newblock in {\em IEEE International Symposium on Biomedical Imaging (ISBI)},
  2018, pp. 734--738.

\bibitem{szegedy2016rethinking}
Christian Szegedy, Vincent Vanhoucke, Sergey Ioffe, et~al.,
\newblock ``Rethinking the inception architecture for computer vision,''
\newblock in {\em Proc. IEEE conference on Computer Vision and Pattern
  Recognition (CVPR)}, 2016, pp. 2818--2826.

\bibitem{parthasarathy2019iap}
A.~Parthasarathy, P.~S.~N. Menon, and M.~K.~C. Nair,
\newblock {\em IAP textbook of pediatrics},
\newblock JP Medical Ltd, 2019.

\bibitem{mao2017least}
X.~Mao, Q.~Li, H.~Xie, et~al.,
\newblock ``Least squares generative adversarial networks,''
\newblock in {\em Proc. IEEE International Conference on Computer Vision
  (ICCV)}, 2017, pp. 2794--2802.

\bibitem{brock2018large}
A.~Brock, J.~Donahue, and K.~Simonyan,
\newblock ``Large scale {GAN} training for high fidelity natural image
  synthesis,''
\newblock in {\em Proc. International Conference on Learning Representations
  (ICLR) arXiv:1809.11096}, 2019.

\end{thebibliography}

\end{document}